# Multipole Expansions of Aggregate Charge: How Far to Go?

Lorin S. Matthews, *Member, IEEE,* Douglas A. Coleman, and Truell W. Hyde, *Member, IEEE*

*Abstract*—**Aggregates immersed in a plasma or radiative environment will have charge distributed over their extended surface. Previous studies have modeled the aggregate charge using the monopole and dipole terms of a multipole expansion, with results indicating that the dipole-dipole interactions play an important role in increasing the aggregation rate and altering the morphology of the resultant aggregates. This study examines the effect that including the quadrupole terms has on the dynamics of aggregates interacting with each other and the confining electric fields in laboratory experiments. Results are compared to modeling aggregates as a collection of point charges located at the center of each spherical monomer comprising the aggregate.**

## I. INTRODUCTION

DUST aggregation plays a crucial role in many astrophysical and terrestrial environments including the early stages of planet formation in protoplanetary disks [1], planetary ring systems [2], cometary tails [3], and atmospheric aerosols [4]. The dynamical behavior of dust in these astrophysical systems can be both interesting and complicated as the dust becomes charged due to the collection of electrons and ions from the local plasma environment or though radiative charging processes. On earth, dust aggregates are present as an unwanted by-product in plasma devices such as fusion reactors and plasma processing systems for silicon wafers [5], [6].

Recently, dust aggregates created in laboratory experiments in GEC rf reference cells have been shown to display interesting dynamical behavior [7]–[11]. For example, interacting aggregates are observed to rotate and change their orientation as they approach each other [8], [9]. Under quiescent conditions, these aggregates tend to be oriented vertically, with their long axis parallel to the sheath electric field and most (though not all) aggregates are observed to rotate about this vertical axis [8], [10]. Such rotations have been attributed to the electrostatic dipole moment created by the charge unevenly distributed over the irregular surface of the aggregate [8], [12].

To properly simulate this behavior, the question arises whether the monopole plus dipole approximation is sufficient to fully capture the dynamical behavior of the aggregates or if higher-order terms must be included. This study examines the effect of including the electrostatic quadrupole moment in the calculation of forces and torques acting on the aggregates and compares the approximation to a model which treats the aggregate charge as a collection of point charges centered at each spherical monomer within the aggregate. The description of the aggregate charging model and the multipole expansion of the electrostatic forces are given in Section II, while Section III compares the relative magnitude of the individual multipole terms and illustrates the effect of including them in dynamical simulations. Conclusions are given in Section IV.

## II. NUMERICAL MODEL

Two different numerical codes are used to model the charging and dynamics of the aggregates. The dynamics code models the interaction of two aggregates, including induced accelerations and rotations, and incorporates all of the external forces present in the system such as gas drag and gravitational and electric fields [13]. The charging code calculates the electron and ion currents to multiple points on the surface of the aggregate and determines the total charge and resulting charge distribution [14].

### A. Charge Distribution on Aggregate

The charge on the surface of an aggregate can be found using orbital motion limited (OML) theory employing a line of sight (LOS) approximation, OML_LOS [14]. The current density due to ions or electrons incident to a point on the surface of a grain is given by

$$J_s = q_s n_s \int_{v_{min}}^{\infty} v_s^3 f(v_s) dv_s \iint \cos \gamma \, d\Omega \qquad (1)$$

with $n_s$ the plasma density of species $s$ very far from the grain, $q_s$ the charge of the incoming plasma particle of mass $m_s$ and temperature $T_s$, $v_s$ the velocity of the incoming plasma particle with a velocity distribution $f(v_s)$, $\gamma$ the angle between the velocity vector and the surface normal, and $\Omega$ the solid angle. The lower limit of integration for the particle velocity is the minimum velocity required for a charged plasma particle to reach a surface point on the aggregate with potential $\Phi$.

A critical point in OML theory is that all positive energy orbits connect back to infinity, and do not originate from

This work was supported in part by the National Science Foundation under Grants PHY-0847127, PHY-1414523, and PHY-0648869.

D. A.Coleman was a REU student at Baylor University, Waco, TX 76798 USA.

L. S. Matthews and T. W. Hyde are with the Center for Astrophysics, Space Physics, and Engineering Research, Baylor University, Waco, TX 76798 USA, (e-mail: lorin_matthews@baylor.edu and truell_hyde@baylor.edu).

another point on the grain [15]. For an irregular aggregate consisting of spherical monomers, the trajectories of incoming plasma particles may be blocked by other monomers in the aggregate, and thus the integral over the angles is approximated by numerically computing the LOS factor, LOS_factor = $\iint \cos \alpha \, d\Omega$ (see [14] for a complete description of this treatment). To accomplish this, the surface of the aggregate is divided into many patches. The LOS factor for each patch is determined by finding the open lines of sight from the center of the patch using 1000 test directions. The current density to each patch is then calculated as a function of the electric potential at the center of that patch due to the charge on all of the patches, including itself. Examples of the charge distribution on the patches and the monomers are shown in Fig. 1.

The charge distribution with respect to the center of mass of the aggregate can be characterized using a multipole expansion. In this case, the total charge on the aggregate, or monopole moment is $q = \sum q_\alpha$, where $\alpha$ is the index of each patch. The dipole moment is given by $\boldsymbol{p} = \sum q_\alpha \boldsymbol{d}_\alpha$, where $\boldsymbol{d}_\alpha$ is the distance of a patch from the center of mass. The elements of the traceless quadrupole moment of the charge distribution, $\overline{\overline{Q}}$, are calculated from

$$Q_{ij} = \sum(3d_{\alpha,i}d_{\alpha,j} - d_\alpha^2 \delta_{ij})q_\alpha. \qquad (2)$$

Using the multipole expansion, the electrostatic potential at a position $\boldsymbol{x}$ from the center of mass of the aggregate is then given by

$$V = \frac{1}{4\pi\epsilon_0}\left(\frac{q}{x} + \frac{\boldsymbol{p} \cdot \boldsymbol{x}}{x^3} + \frac{x^T \overline{\overline{Q}} x}{2x^5}\right). \qquad (3)$$

A more accurate representation of the charge distribution is to assume the total charge on each monomer acts as a point charge located at the center of the monomer, given by the vector $\boldsymbol{r}$ with respect to the aggregate center of mass. In this case, the equation for the potential is given by

$$V = \frac{1}{4\pi\epsilon_0}\sum_\beta \frac{q_\beta}{|\boldsymbol{x} - \boldsymbol{r}_\beta|} \qquad (4)$$

where the index $\beta$ runs over the total number of monomers in the aggregate.

### B. Multipole Expansion of Forces and Torques

Aggregates in a RF plasma levitate in the sheath region near the charged electrode. The vertical electric field in this region, to a good approximation, can be considered linear [16]. Designating the electric field at the levitation height as $E_0$, the electric field can be written as

$$E_z = E_0 + E' z \qquad (5)$$

where $E'$ is the gradient in the vertical direction. Most experiments also include a mechanism such as a circular depression in or glass box placed on the lower electrode to confine the dust particles in the radial direction. Numerous measurements have found the horizontal electric field near the

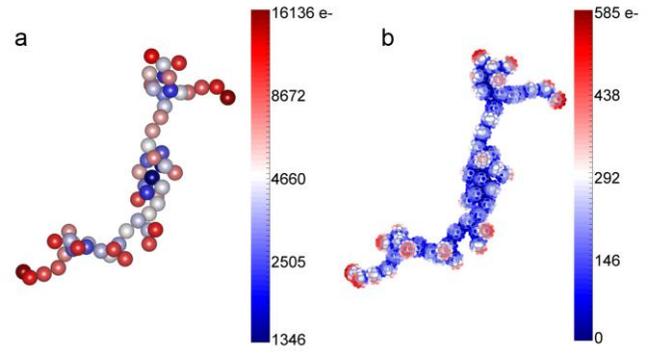

Figure 1. (Color online) Charge distribution on an aggregate. Total number of elementary charges a) on each monomer and b) on each patch.

center of the confining region to be radially symmetric [17], [18] with

$$E_r = kr \qquad (6)$$

where $k$ is a constant which sets the strength of the radial confinement.

The multipole approximation of forces and torques due to electrostatic interactions can be expanded to include terms up to the dipole-quadrupole interactions. The electric field of a charged aggregate at a point $\boldsymbol{x}$ from its center of mass is determined from

$$\boldsymbol{E}(\boldsymbol{x}) = \frac{1}{4\pi\epsilon_0}\left[\frac{q\hat{\boldsymbol{x}}}{x^2} + \frac{3\hat{\boldsymbol{x}}(\boldsymbol{p}\cdot\hat{\boldsymbol{x}}) - \boldsymbol{p}}{x^3} + \frac{1}{x}\left(\frac{5}{2x^2}\left(\boldsymbol{x}^T\overline{\overline{Q}}\boldsymbol{x}\right)\boldsymbol{x} - \overline{\overline{Q}}\boldsymbol{x}\right) + \cdots\right]$$

$$= \boldsymbol{E}_{\text{mon}} + \boldsymbol{E}_{\text{dip}} + \boldsymbol{E}_{\text{quad}} + \cdots \qquad (7)$$

where $x = |\boldsymbol{x}|$.

The force on an aggregate with charge $q$ and dipole moment $\boldsymbol{p}$ which is in a non-homogeneous electric field is given by

$$\boldsymbol{F} = q\boldsymbol{E} + \overline{\overline{G}}\boldsymbol{p}, \qquad (8)$$

where $\overline{\overline{G}}$ is the electric field gradient, $G_{ij} = \nabla_i E_j$. If the non-homogeneous electric field is due to another aggregate with its charge characterized by a multipole expansion up to the quadrupole term, the gradient is given by

$$G_{ij} = \frac{-1}{8\pi\epsilon_0 x^7}[30x_i x_j \boldsymbol{p} \cdot \boldsymbol{x}$$
$$+ 35x_i x_j \boldsymbol{x}^T \overline{\overline{Q}} \boldsymbol{x} - 10x_i \Sigma_k Q_{jk} x_k - 10x_j \Sigma_k Q_{ik} x_k$$
$$+ r^2\left(2Q_{ij} - 6p_i x_j - 6p_j x_i + 6qx_i x_j\right)$$
$$+ \delta_{ij}(5\boldsymbol{x}^T\overline{\overline{Q}}\boldsymbol{x} + 6x^2\boldsymbol{p} \cdot \boldsymbol{x} + 2qx^4)]. \qquad (9)$$

Higher order moments of the electric force expansion in (8) and (9) are given in [19].

The torque on a dipole in an electric field is calculated as $N_i^{\text{dip}} = \sum_{j,k} \mathcal{E}_{ijk} p_j E_k$, where $\mathcal{E}_{ijk}$ is the Levi-Civita pseudo-tensor. In addition, the torque on a quadrupole by the electric field gradient is given by [20]

$$N_i^{\text{quad}} = \tfrac{1}{3} \sum_{jkl} \mathcal{E}_{ijk} Q_{jl} G_{kl}. \qquad (10)$$

Considering the more accurate representation of the charge distribution using the charge on each monomer, the force on each monomer in aggregate 1 due to the charge on all monomers in aggregate 2 is calculated from

$$\boldsymbol{F}_\alpha = \sum_\beta k q_\alpha q_\beta / r_{\alpha\beta}^2 \hat{r}_{\alpha\beta} \qquad (11)$$

where the indices $\alpha$ and $\beta$ run over all the monomers in aggregate 1 and aggregate 2, respectively, and $r_{\alpha\beta}$ is the distance between two monomers. The net force on aggregate 1 is then given by $F_1 = \sum_\alpha F_\alpha$, with a net torque given by $\boldsymbol{N}_1 = \sum_\alpha \boldsymbol{r}_\alpha \times \boldsymbol{F}_\alpha$, with similar expressions for aggregate 2.

## III. RESULTS

Multipole moments were calculated for charged aggregates to determine the relative contributions of the monopole, dipole, and quadrupole terms. The accuracy of these approximations was gauged by comparison with the results obtained by modeling the aggregate as a collection of discrete point charges centered at each monomer.

### A. Characterization of multipole moments

A collection of aggregates consisting of 2 to 2000 monomers (see [14] for details) was charged using OML_LOS. The monopole, dipole, and quadrupole moments for each aggregate were calculated using the location and magnitude of the charge on each patch. Fig. 2a shows the charge on the aggregates, which increases nearly linearly with the aggregate radius $R$, defined as the greatest extent of the aggregate from its center of mass. Fig. 2b shows the normalized multipole moments. The magnitude of the dipole moment is normalized by dividing by $qR$, while the magnitude of the quadrupole tensor, defined in this case as the root-mean-square of the eigenvalues, is normalized by $qR^2$. While there is a large spread in the multipole values, in general the magnitude of the dipole moment is approximately $10^{-2} qR$ while the magnitude of the quadrupole moments is about $10^{-1} qR^2$. These normalized values will be used for comparing the relative strength of the multipole terms.

### B. Potential field lines

In Fig. 3, the electrostatic potential is shown for three different orders of the multipole field expansion, plus the calculation of the electric potential using the charge $q_\beta$ centered at each monomer within the aggregate. In this case, the potential due to the charges on the individual monomers, indicated by the thin black lines, is assumed to be representative of the actual potential. The quadrupole expansion, indicated by the dashed lines, gives a good fit to the potential at distances $\gtrsim 2R$, as expected. However, it significantly deviates from the actual potential for $x < R$. The dipole approximation (dash-dot line) does not differ significantly from the monopole approximation (dotted line), but these approximations tend to only be accurate for distances greater than $2R$.

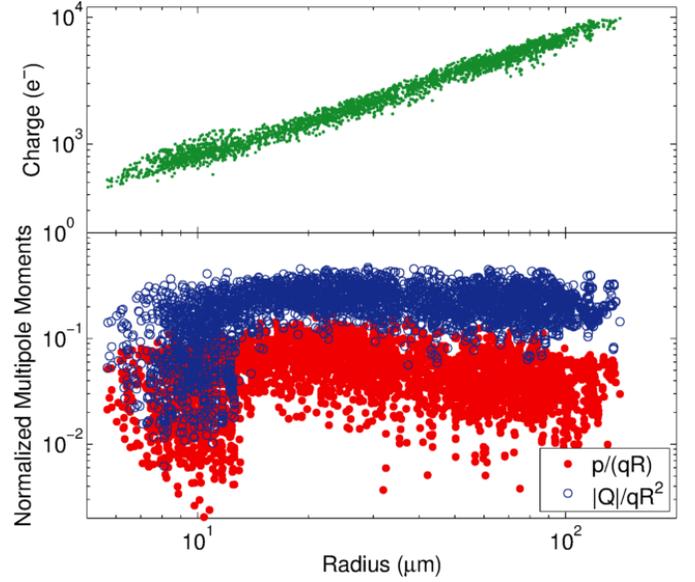

Figure 2. (Color online) Charge (a) and normalized multipole moments (b) as a function of aggregate radius. The number of monomers in the aggregates range from $N = 2$ to $N = 2000$.

### C. Dynamics of Aggregate in Linear Electric Field

It is helpful to estimate the magnitude of the relative contributions to the force and torque to determine which terms may be safely neglected in the multipole expansion. In a linear electric field such as those found in a discharge plasma, an aggregate will experience an acceleration and a torque. Considering only electric fields that can be described by (5) and (6), the gradient of the electric field will only have diagonal terms. As an example, combining (5) and (8) the force is given by

$$F_z = q E_0 + E' p_z. \qquad (12)$$

In a typical argon discharge plasma, the magnitude of the vertical electric field is of order O($10^3$ V/m), while $E'$ is of order O($10^6$ V/m$^2$) [16]. In this case, one might expect the dipole contribution to be important when the second term is a few percent of the first term. Taking the ratio of the dipole term to the monopole term gives

$$\frac{E' p_z}{q E_0} \approx \frac{10^{-1} q R E'}{q E_0} \approx \frac{(10^{-2})(10^{-5})(10^6)}{10^3} = 10^{-4} \qquad (13)$$

where $R \sim 10^{-5}$ m has been used for the approximate size of a large aggregate [8], [10]. Thus the dipole contribution to the acceleration is very small and can be neglected. A similar result is found for the radial direction.

The dipole moment does lead to a torque acting on the aggregate, however. Note that in a constant electric field the dipole torque $N_i^{\text{dip}} = \sum_{j,k} \mathcal{E}_{ijk} p_j E_k$ is exactly equal to the torque due to the electric force $q\boldsymbol{E}$ acting at the center of charge of the aggregate ($COQ = \frac{1}{q}\Sigma_\alpha q_\alpha \mathbf{d}_\alpha$), which is displaced from the center of mass. The deviation from a constant electric field is very small for an aggregate in the sheath of an argon plasma, as discussed above, so the monopole term for the aggregate can be used to determine both the acceleration and the rotation with relative accuracy in this case. The ratio of the quadrupole torque, given by (10), to the dipole torque is

$$\frac{\frac{1}{2}|Q|E'}{|p|E_0} = \frac{\frac{1}{3}10^{-1}qR^2E'}{10^{-2}qRE_0} = \frac{10RE'}{3E_0} \approx \frac{10(10^{-5})(10^6)}{3(10^3)} = 10^{-2}. \quad (14)$$

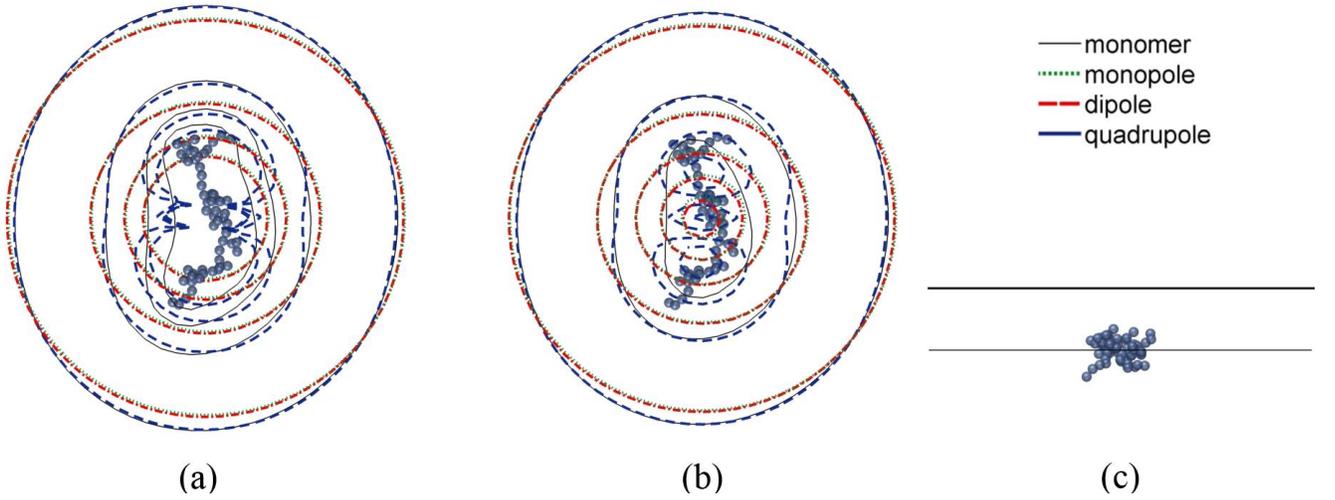

Figure 3. (Color online) Lines of constant electric potential for a) the midplane and b) a plane just above the aggregate. c) Side view showing location of the planes. The legend indicates the terms used in the expansion to calculate the potential. For this particular aggregate, $p = 0.025qR$ and $Q = 0.42qR^2$.

Thus the quadrupole torque provides a small correction.

As an illustration, the orientation of an aggregate suspended in the sheath of an argon rf discharge is shown in Fig. 4. The forces acting on the aggregate include gravity and confining electric fields of the form given in (12) and (13). Gas drag is added to the simulation to allow the aggregate to reach its equilibrium configuration. Arrows indicate the direction of the aggregate dipole in three different cases: 1) the torque caused by electric field acting at the center of charge (monopole term only), 2) the dipole torque, and 3) the dipole plus the quadrupole torque. The dipole axes in the first two cases are almost perfectly aligned with the vertical electric field, while in the third case the dipole axis is aligned to within a tenth of a degree. This slight misalignment causes the aggregate to rotate about the vertical axes at a constant rate, behavior which has been observed in the lab [8], [10].

### D. Dynamics of two interacting aggregates

Calculating the interactions between two aggregates separated by a distance $\mathbf{x}$ requires comparison of the magnitudes of the dipole-dipole and dipole-quadrupole terms. Expanding the force in (8) in terms of the multipole moments yields

$$\mathbf{F}_{12} = q_1\big(\mathbf{E}_{2,mon} + \mathbf{E}_{2,dip} + \mathbf{E}_{2,quad}\big) \\ + \big(\overline{\overline{\mathbf{G}}}_{2,mon} + \overline{\overline{\mathbf{G}}}_{2,dip} + \overline{\overline{\mathbf{G}}}_{2,quad}\big)\mathbf{p}_1 \quad (15)$$

where the subscripts one and two refer to the first and second aggregate. We would like to compare the magnitude of each successive term to the leading monopole-monopole term, $|q_1E_{mon}| \propto q_1q_2/x^2$. Thus

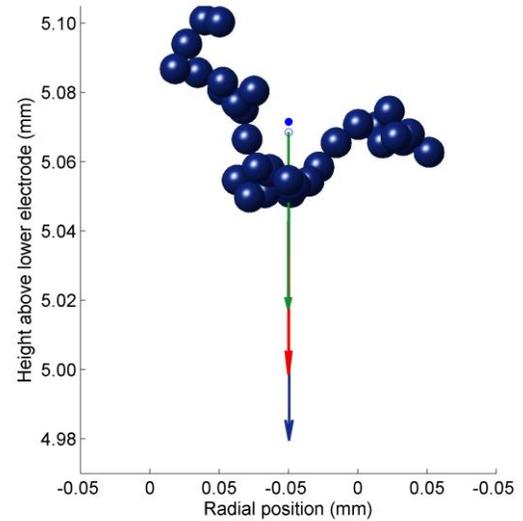

Figure 4. (Color online) Orientation of an aggregate levitating in the sheath of an rf discharge. The closed circle indicates the location of the center of charge, while the open circle indicates the location of the center of mass. The direction of the dipole moment for the three cases described in the text are shown, with the upper green arrow representing the monopole torque acting at the center of charge, the middle red arrow the dipole torque, and the lower blue arrow the quadrupole plus dipole torque.

$$\frac{|q_1E_{2,dip}|}{|q_1E_{2,mon}|} \approx \frac{p_2}{q_2x} \approx \frac{10^{-2}q_2R_2}{q_2x} = \frac{R_2}{10^2x} = \frac{1}{10^2\xi}, \qquad (16a)$$

$$\frac{|q_1E_{2,quad}|}{|q_1E_{2,mon}|} \approx \frac{Q_2}{q_2x^2} \approx \frac{10^{-1}q_2R_2^2}{q_2x^2} = \frac{R_2^2}{10x^2} = \frac{1}{10\xi^2}. \qquad (16b)$$

Plots of these values as a function of $\xi$, the ratio of the distance from the aggregate to the aggregate radius, are shown in Figure 5a. As the multipole approximation is not valid for values of $\xi$ less than one, it is clearly evident that dipole contribution to the acceleration will always be small, although the quadrupole contribution becomes increasingly important as $\xi$ decreases. A similar result is found for the second set of terms in (15), using the magnitude of $p_1$ and gradient of the electric field $\vec{G}_2$.

Comparison of the relative contribution to the torques is more interesting, however. The leading term for $N^{dip}$ is $p_1E_{2,mon} \approx 10^{-2}q_1R_1q_2/x^2$, while the leading term for $N^{quad}$ is $10^{-1}q_1R_1^2q_2/x^3$. Thus

$$\frac{N^{quad}}{N^{dip}} = \frac{10R_1}{x} = \frac{10}{\xi} \qquad (17)$$

and for intermediate separation distances $1 < \xi < 10$ the quadrupole torque is the dominant term, as illustrated in Fig. 5b.

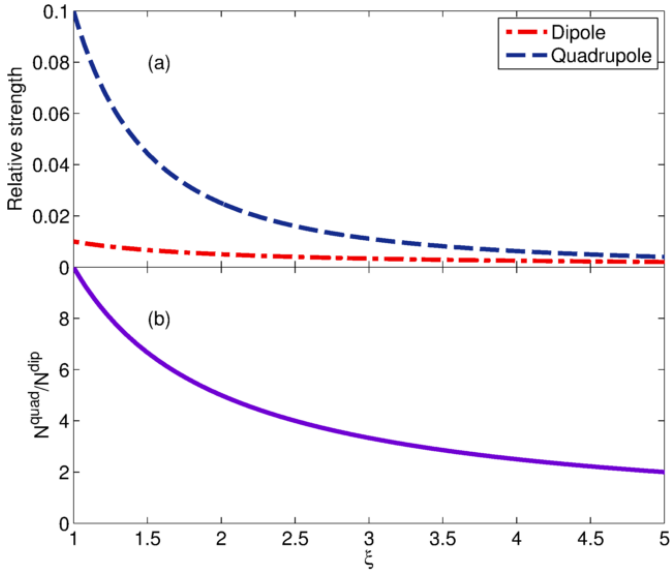

Figure 5. (Color online) a) Ratio of the dipole and quadrupole contributions to the electrostatic force relative to the monopole contribution and b) ratio of the quadrupole torque to the dipole torque as a function of the normalized distance between aggregates.

As an example of the differences that arise during a two-particle interaction, snapshots of images are shown at equal time intervals in Fig. 6 for calculations using the different electric field approximations. In Fig 6a, the electric field and torques are calculated using the charge on each monomer. The rotations of the two aggregate are considerable, and the aggregates ultimately move apart without touching. Using the quadrupole moment of the aggregates, the trajectory is almost unaltered. However, the differing rotations and accelerations in this case at close proximity allow the two aggregates to collide and stick (Fig. 6b). Fig. 6c shows the interaction considering only the dipole terms for the two electric fields. In this case, the target aggregate rotates in the opposite sense, and the aggregate approaching from below is repelled. A simulation using only the monopole terms shows a trajectory almost identical to the dipole case, but the aggregate orientations remain fixed.

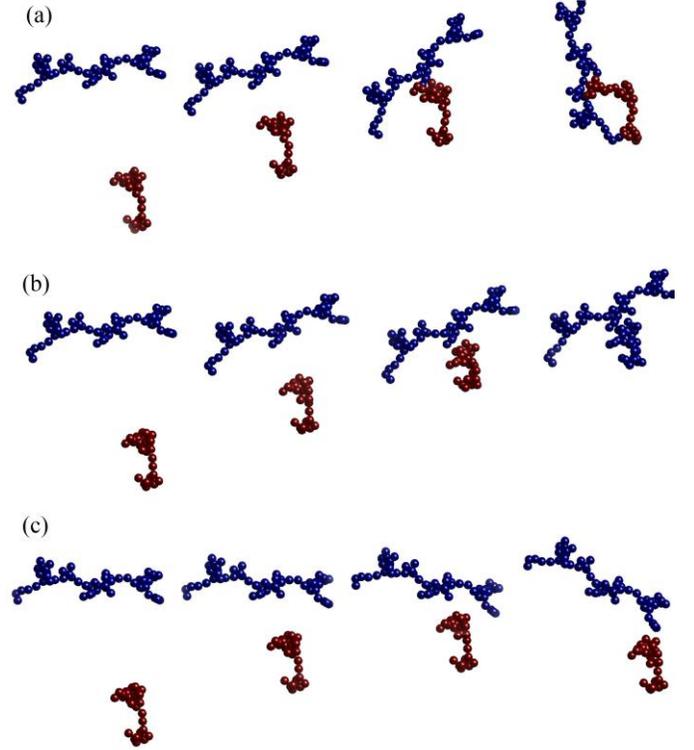

Figure 6. (Color online) Trajectories of two interacting aggregates. The frames in each row are taken at equal time intervals, with the camera fixed on the center of mass of the upper aggregate. Calculation using the potential due to a) the charge on each monomer, b) the quadrupole moments of the aggregates, and c) the dipole moments of the aggregates. The trajectory with only the monopole interactions is almost identical to this case, with no rotations of the aggregates.

## IV. CONCLUDING REMARKS

The dynamics of charged aggregates has been examined by including various terms of the multipole expansion of the electric potential. The charge on the aggregate tends to be greatest at the extremities of the aggregate, and the dipole approximation alone is not suitable for this charge arrangement, which is better represented by an expansion up to the quadrupole term. In the case of an aggregate in a uniform electric field, or an electric field which varies slowly over distances comparable to the aggregate size, the acceleration and torque can be accurately treated using just the monopole approximation, if the electric force is assumed to act at the aggregate's center of charge. For slowly varying electric fields, including the terms up to the quadrupole torque gives a small improvement, but the dynamical behavior can be quite different, as shown in Fig. 4.

The electrostatic interaction between two charged grains requires higher orders of the multipole expansion. The torques acting on the aggregates are dominated by the quadrupole interaction, which can be several times greater than the torques from the dipole moment. However, in the case of collisions between aggregates, great accuracy is needed at small distances, where the multipole expansion breaks down. At short distances of less than a few aggregate radii it is more accurate to calculate the electrostatic interactions by treating the aggregate as a collection of fixed point charges centered at each monomer within the aggregate.

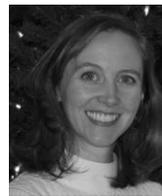

**Lorin S. Matthews** received the B.S. and the Ph.D. degrees in physics from Baylor University in Waco, TX, in 1994 and 1998, respectively. She is an Associate Professor in the Physics Department at Baylor University and Associate Director of the Center for Astrophysics, Space Physics, and Engineering Research (CASPER). Previously, she worked at Raytheon Aircraft Integration Systems as the Lead Vibroacoustics Engineer on NASAs SOFIA (Stratospheric Observatory for Infrared Astronomy) project.

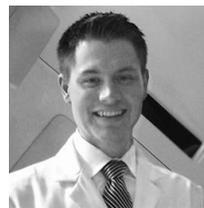

**Douglas A. Coleman** received a B.S. in physics from Grand Valley State University in 2011 and a Doctorate of Medical Physics (D.M.P.) from Vanderbilt University in 2015. He is now pursuing ABR board certification while practicing clinical medical physics in radiation oncology.

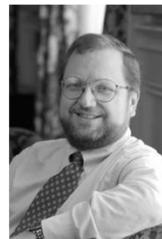

**Truell W. Hyde** was born in Lubbock, Texas. He received the B.S. in physics and mathematics from Southern Nazarene University in 1978 and the Ph.D. in theoretical physics from Baylor University in 1988. He is currently at Baylor University where he is the Director of the Center for Astrophysics, Space Physics & Engineering Research (CASPER), a Professor of physics and the Vice Provost for Research for the University. His research interests include space physics, shock physics and waves and nonlinear phenomena in complex (dusty) plasmas.